\documentclass[twocolumn,showpacs,superscriptaddress,prl]{revtex4}
\usepackage{graphicx}
\usepackage{amsmath}
\usepackage{amsthm}
\usepackage{amssymb}
\usepackage{latexsym}
\usepackage{array}
\usepackage{float}
\usepackage{amsfonts}
\usepackage{mathrsfs}
\usepackage{verbatim}

\usepackage{color}

\newcommand{\bra}[1]{\left<#1\right|}
\newcommand{\ket}[1]{\left|#1\right>}
\newcommand{\abs}[1]{\left|#1\right|}

\begin{document}

\title{Optimal Usage of Quantum Random Access Memory in Quantum Machine Learning}

\author{Jeongho Bang}
\affiliation{School of Computational Sciences, Korea Institute for Advanced Study, Seoul 02455, Korea}
\affiliation{Institute of Theoretical Physics and Astrophysics, University of Gda\'{n}sk, 80-952 Gda\'{n}sk, Poland}

\author{Arijit Dutta}
\affiliation{School of Computational Sciences, Korea Institute for Advanced Study, Seoul 02455, Korea}

\author{Seung-Woo Lee}\email{swleego@gmail.com}
\affiliation{Quantum Universe Center, Korea Institute for Advanced Study, Seoul 02455, Korea}

\author{Jaewan Kim}\email{jawan@kias.re.kr}
\affiliation{School of Computational Sciences, Korea Institute for Advanced Study, Seoul 02455, Korea}

\received{\today}

\begin{abstract}
By considering an unreliable oracle in a query-based model of quantum learning, we present a tradeoff relation between the oracle's reliability and the reusability of quantum state of the input data. The tradeoff relation manifests as the fundamental upper bound on the reusability. This limitation on the reusability would increase the quantum access to the input data, i.e., the usage of quantum random access memory (qRAM), repeating the preparation of a superposition of ``big'' input data on the query failure. However, it is found that, a learner can obtain a correct answer even from an unreliable oracle without any additional usage of qRAM---{\em i.e., the complexity of qRAM query does not increase even with an unreliable oracle.} This is enabled by repeatedly cycling the quantum state of the input data to the upper bound on the reusability.
\end{abstract} 


\maketitle

{\em Introduction.}---Quantum machine learning (QML) is a rapidly growing research field currently. A primary issues in QML is the identification of a quantum advantage over the classical counterparts~\cite{Schuld15,Biamonte17,Ciliberto18}. A recent proposal of the quantum support vector machine (QSVM)~\cite{Rebentrost14}, providing an exponential speed-up in a classification task, can be considered as a paradigmatic achievement in this direction. Currently, the QSVM (and other variant QML proposals~\cite{Rebentrost14,Lloyd14,Schuld16,Kerenidis16,Liu18}) provides a standardized approach to achieve quantum speed-up---that is, to use the quantum state provided that a set of input data are superposed in a weighted distribution.

However, unclear aspects still exist in QML. In particular, whether the quantum advantage remains significant even when the cost to access ``big'' input data is considered needs to be clarified; i.e., whether classical input data can be transformed to a quantum superposition~\cite{Aaronson15,Zhao18}. In theory, at least, the quantum random access memory (qRAM) can accomplish the aforementioned task~\cite{Giovannetti08-1,Giovannetti08-2}, even though its realization is far from trivial~\cite{Arunachalam15}. Subsequently, a question arises as to whether it is possible to reduce the qRAM query by reusing the quantum state of the input data that has been initialized once. The reuse of quantum data is limited because the information-extraction causes the state-disturbance, and/or due to the no-cloning theorem~\cite{Wootters82}, contrary to the classical machine learning that has no limitation in reusing the data~\cite{Custers16}. However, an original state can be retrieved using weak measurements with non-zero probability~\cite{Koashi99, Cheong12}. Hence, it is importnat to explore whether the reuse or recycle of the quantum state of the input data is possible, the quantum limit on the reusability, and whether it offers any advantage in QML.

The oracle's reliability also affects the learning performance significantly~\cite{Iwama06}. The effects by the unreliable oracle with missed answer or evasive answer (e.g. ``I do not know'') have been studied and shown to be tolerable in query-based models of classical learning~\cite{Angluin94,Simon04}. In particular, the learner can be polynomially dominated by a failure query rate of less than $\frac{1}{2}$~\cite{Angluin94,Simon04}. Such results were also drawn in QML~\cite{Iwama06,Ciliberto18}. Furthermore, it was claimed that some quantum advantages are achievable with noisy oracles~\cite{Cross15,Diego17}. However, the effects of the oracle's reliability on the complexity of qRAM query have not been studied in QML, even though recent speed-ups of QML hinge crucially on the low qRAM queries.


Herein, by casting a query-based model of quantum leaning with an unreliable oracle, we explore the fundamental limit on the reusability of the quantum state of the input data quantitatively. In particular, we present a tradeoff relation between the oracle's reliability and the reusability of quantum state of the input data. The tradeoff relation indicates that the more reliable the oracle, the lower is the reusability. It also manifests the fundamental upper bound on the reusability for given oracle reliability. Such a limited reusability of the quantum input data would impose the additional usage of qRAM, thus repeating the quantum access to the input data with query failure. However, it is found that the learner can, in principle, arrive at the correct answer with a single run of qRAM~\footnote{Throughout the work, it is assumed that the complexity of qRAM query is primarily related to the (quantum) access to the input data to initialize a quantum superposition of the input data~\cite{Aaronson15,Zhao18}.}, repeatedly cycling the quantum state of the input data to the upper bound of the reusability. This result implies that, if the traveling cost of the input data is neglected, an incomplete-oracle learner has the same complexity of qRAM query as that of a complete-oracle learning.

\begin{figure}[t]
\centering
\includegraphics[width=0.46\textwidth]{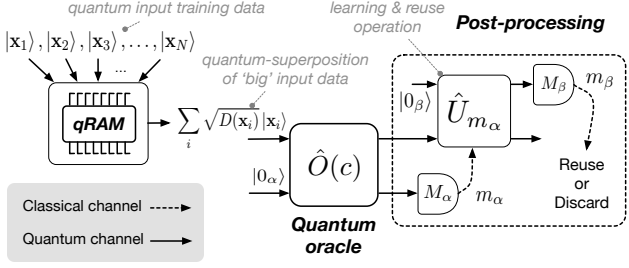}
\caption{\label{fig:basic_archit} A schematic picture of the query-based QML model. Firstly, the quantum random access memory (qRAM) initializes a superposition of ``big'' (say $N$) input data. Then, the quantum oracle $\hat{O}(c)$ is employed. For a general scenario, $\hat{O}(c)$ is assumed to be unreliable, yielding an incorrect answer with a non-zero probability. The last post-processing block (dashed box) is responsible for the learning and the reuse of the superposed state initialized by qRAM (see the main text).}
\end{figure}

{\em Model.}---Typically, machine learning is often formulated as an identification of a function $c$ (referred to as a ``concept'' in the language of machine learning); it maps the input data $\mathbf{x} \in \{0,1\}^n$ (in arbitrary $n$-bit strings) to the target $c(\mathbf{x}) \in \{0, 1\}$---i.e., a task of classification~\cite{Langley95}. In contrast to classical machine learning, QML employs a set of quantum training data, i.e., $\ket{\mathbf{x}}$ and $\ket{c(\mathbf{x})} \in \{\ket{0}, \ket{1}\}$. Hence, we design a query-based QML model, as shown in Fig.~\ref{fig:basic_archit}.

Our model is roughly composed of three blocks. The first block is the initialization of a quantum superposition of the `big' input data $\ket{\mathbf{x}_i}$ ($i=1,2,\ldots,N$). At least in theory, this task can be accomplished by casting the qRAM~\cite{Giovannetti08-1}; more specifically, qRAM allows for the data to be read (or to be written) from arbitrary $i$-th memory cells \footnote{Here, the memory arrays are either classical or quantum depending on the accessing type of qRAM.} and creates the superposition of all $N$ input data, denoted hereinafter as 
\begin{eqnarray}
\ket{\psi_0} = \sum_{i=1}^N\sqrt{D_i} \ket{\mathbf{x}_i},
\label{eq:qid}
\end{eqnarray}
where $D_i$ is a probability distribution of memory cells.

We then consider the quantum learning oracle $\hat{O}(c)$ that is assumed to be unreliable, yielding an incorrect answer $\ket{c(\mathbf{x}_i) \oplus 1}$ with a certain probability \cite{Cross15}. The oracle operation is defined as
\begin{eqnarray} \ket{\psi_0}\ket{0_\alpha} \xrightarrow{\hat{O}(c)} &\displaystyle \sum_{i=1}^N&\sqrt{D(\mathbf{x}_i)} \Big( \sqrt{\lambda_{+}} \ket{\mathbf{x}_i}\ket{c(\mathbf{x}_i)} \nonumber \\
    && ~~~~~ + \sqrt{\lambda_{-}} \ket{\mathbf{x}_i}\ket{c(\mathbf{x}_i) \oplus 1}\Big),
\label{eq:oracle_op}
\end{eqnarray}
where 
\begin{eqnarray}
\lambda_\pm = \frac{1 \pm {\cal L}}{2}
\end{eqnarray}
is the qubit state of the oracle-answer register. Here, we define the {\em oracle's reliability} with the factor ${\cal L} \in [0,1]$; for example, $\hat{O}(c)$ is perfectly reliable when ${\cal L}=1$, but is less reliable when ${\cal L} < 1$. For the case when ${\cal L}=0$, the oracle $\hat{O}(c)$ produces a completely random answer, yielding no information. We clarify that the queries to qRAM and $\hat{O}(c)$ are distinct; the qRAM query is engaged as the process of initializing a superposition [as in Eq.~(\ref{eq:qid})] of the input data, while the oracle $\hat{O}(c)$ is queried about the legitimate learning output for the (superposed) inputs.

The last block is for post-processing, i.e., learning and the reuse of $\ket{\psi_0}$ in Eq.~(\ref{eq:qid}). Prior to those processes, the oracle's answer, or equivalently the learning information, should be identified within this block. Thus, a projection measurement $M_\alpha$ is assumed first to yield (the information of) the oracles's answer, followed by sequential operations denoted by $\hat{U}_{m_\alpha}$. The information for learning can be extracted from the measurement outcome $m_\alpha$. After the measurement, $m_\alpha$ is delivered to and utilized by $\hat{U}_{m_\alpha}$ for the learning and/or reuse process. To recycle $\ket{\psi_0}$, the operation $\hat{U}_{m_\alpha}$ is manipulated according to the outcome $m_\alpha$ and is applied to the output state of $\hat{O}(c)$. Another projection measurement $M_\beta$ is performed after $\hat{U}_{m_\alpha}$. Here, we define {\em the reusability}, denoted by ${\cal R}$, in terms of the overall probability of attaining $\ket{\psi_0}$ after the measurement $M_\beta$.

It is worth noting that our model is conceptually equivalent to the conventional query-based model of learning \cite{Lyubashevsky05}, by which the best speed-up is polynomial \cite{Servedio04,Kothari14}. However, employing such a model is sufficient to derive a quantitative relation between the oracle's reliability and the reusability of $\ket{\psi_0}$.

{\em Tradeoff relation.}---We herein present a tradeoff relation between the reliability ${\cal L}$ of the oracle $\hat{O}(c)$ and the reusability ${\cal R}$ of the quantum state $\ket{\psi_0}$ of the input data. For convenience in calculations, we rewrite Eq.~(\ref{eq:qid}) as the following form:
\begin{eqnarray}
\ket{\psi_0} = \sum_{i=1}^N \sqrt{D(\mathbf{x}_i)}\ket{\mathbf{x}_i} = \sum_{\tau=0,1} \sqrt{\xi_\tau} \ket{X_\tau},
\label{eq:state_X}
\end{eqnarray}
where 
\begin{eqnarray}
\ket{X_\tau}=\sum_{\mathbf{x}_i \in X_\tau} \sqrt{\frac{D(\mathbf{x}_i)}{\xi_\tau}} \ket{\mathbf{x}_i}~\text{and}~\xi_\tau = \sum_{\mathbf{x}_i \in X_\tau} D(\mathbf{x}_i). 
\end{eqnarray}
Here, $X_\tau \subset \{ \mathbf{x}_i : i=1,\ldots,N\}$ denotes a set of $\mathbf{x}_i$, satisfying $c(\mathbf{x}_i) = \tau$ and $\cup_{\tau=0,1} X_\tau = \{ \mathbf{x}_i : i=1,\ldots,N\}$. 

Subsequently, we introduce a set of Kraus operators $\hat{A}_{m_\alpha}$ ($m_\alpha = 0, 1$), defined by the combination of $\hat{O}(c)$ and $M_\alpha$. By adopting a fixed form of $\hat{O}(c)$ as
\begin{eqnarray}
\hat{O}(c)=\sum_{\tau=0,1}&\ket{X_\tau}\bra{X_\tau}&\Big( \sqrt{\lambda_{+}}\ket{c=\tau}\bra{0} \nonumber \\
 && + \sqrt{\lambda_{-}}\ket{c=\tau \oplus 1}\bra{0}\Big),
\end{eqnarray}
we can characterize $\hat{A}_{m_\alpha}$ such that
\begin{eqnarray}
\hat{A}_{0} &=& \sqrt{\lambda_{+}} \ket{X_0}\bra{X_0} + \sqrt{\lambda_{-}} \ket{X_1}\bra{X_1}, \nonumber \\
\hat{A}_{1} &=& \sqrt{\lambda_{-}} \ket{X_0}\bra{X_0} + \sqrt{\lambda_{+}} \ket{X_1}\bra{X_1},
\end{eqnarray}
with the eigenvalues $\sqrt{\lambda_\pm}$. The process of extracting the learning information is subsequently expressed as follows:
\begin{eqnarray}
\hat{A}_{m_\alpha}\ket{\psi_0} = \sqrt{P_{m_\alpha}}\ket{\varphi_{m_\alpha}}~(m_\alpha = 0,1),
\label{eq:Apsi0}
\end{eqnarray}
where $P_{m_\alpha}$ is given as
\begin{eqnarray}
P_0 = \frac{1 + \left(\xi_0 - \xi_1\right){\cal L}}{2},~\text{and}~P_1 = \frac{1 - \left(\xi_0 - \xi_1\right){\cal L}}{2}.
\label{eq:P_c}
\end{eqnarray}

We also define an operator $\hat{R}^{(m_\alpha)}$ for the reuse process as the combination of $\hat{U}_{m_\alpha}$ and $M_\beta$. The reuse process is subsequently expressed as
\begin{eqnarray}
\hat{R}^{(m_\alpha)} \hat{A}_{m_\alpha} \ket{\psi_0} = \sqrt{\eta^{(m_\alpha)}} \ket{\psi_0},
\end{eqnarray}
where $\eta^{(m_\alpha)}$ is a non-zero complex number. Because $\hat{\openone} - \hat{R}^{(m_\alpha)}{}^\dagger \hat{R}^{(m_\alpha)}$ is positive semidefinite,
\begin{eqnarray}
\sup_{\ket{\chi}} \bra{\chi}\hat{R}^{(m_\alpha)}{}^\dagger\hat{R}^{(m_\alpha)}\ket{\chi} \le 1,
\label{eq:positive-semi}
\end{eqnarray}
for arbitrary normalized states $\ket{\chi}$. Meanwhile, for the state $\ket{\varphi_{m_\alpha}}=\frac{\hat{A}_{m_\alpha}\ket{\psi_0}}{\sqrt{P_{m_\alpha}}}$ \cite{Koashi99},
\begin{eqnarray}
&& \sup_{\ket{\chi}} \bra{\chi}\hat{R}^{(m_\alpha)}{}^\dagger\hat{R}^{(m_\alpha)}\ket{\chi} \nonumber \\
&& ~~~~ \ge \sup_{\ket{\varphi_{m_\alpha}}} \bra{\varphi_{m_\alpha}}\hat{R}^{(m_\alpha)}{}^\dagger\hat{R}^{(m_\alpha)}\ket{\varphi_{m_\alpha}} \nonumber \\
&& ~~~~ = \sup_{\ket{\psi_0}} \frac{\bra{\psi_0}\hat{A}_{m_\alpha}^\dagger\hat{R}^{(m_\alpha)}{}^\dagger\hat{R}^{(m_\alpha)}\hat{A}_{m_\alpha}\ket{\psi_0}}{P_{m_\alpha}} \nonumber \\
&& ~~~~ = \inf_{\ket{\psi_0}}\frac{\eta^{(m_\alpha)}}{P_{m_\alpha}},
\end{eqnarray}
and by Eq.~(\ref{eq:positive-semi}), we obtain $\eta^{(m_\alpha)} \le \inf_{\ket{\psi_0}}{P_{m_\alpha}}$. Subsequently, from Eq.~(\ref{eq:P_c}), we can verify that $\inf_{\ket{\psi_0}}{P_{m_\alpha}} = \lambda_{-}$ when $\xi_0 - \xi_1 = -1$ for $m_\alpha=0$ and when $\xi_0 - \xi_1 = 1$ for $m_\alpha=1$. It is worth noting that generally the initial state can be written with an arbitrary orthonormal bases and coefficients according to the choice of $M_\alpha$. Thus, for all $m_\alpha=0,1$, we can obtain $\eta^{(m_\alpha)} \le \lambda_{-}$. Then, the probability of attaining the reusable $\ket{\psi_0}$, for an $m_\alpha \in \{0, 1\}$, is bounded as
\begin{eqnarray}
\abs{\bra{\psi_0}\hat{R}^{(m_\alpha)}\ket{\varphi_{m_\alpha}}}^2 = \frac{\eta^{(m_\alpha)}}{P_{m_\alpha}} \le \frac{\lambda_{-}}{P_{m_\alpha}}.
\label{eq:P_reuse_malpha}
\end{eqnarray}

We can finally obtain the overall success probability of the reuse, i.e., the reusability, as 
\begin{eqnarray}
{\cal R} = \sum_{m_\alpha=0,1} P_{m_\alpha} \abs{\bra{\psi_0}\hat{R}^{(m_\alpha)}\ket{\varphi_{m_\alpha}}}^2 \le 1-{\cal L}.
\label{eq:tradeoff_LR}
\end{eqnarray}
This clearly shows that ${\cal R}$ is inversely correlated with and limited by ${\cal L}$; i.e., that of a tight tradeoff relation between the reusability and the oracle's reliability. Note that our proof is valid for arbitrary $\hat{U}_{m_\alpha}$ and $M_\beta$. This result is in agreement with the theorem made in the information-theoretic perspectives~\cite{Cheong12}. 

The tradeoff relation in Eq.~(\ref{eq:tradeoff_LR}) manifests the fundamental limit on the reusability of the quantum state of the input data in QML. The average reusable number is given by
\begin{eqnarray}
\overline{n} = \sum_{n=0}^{\infty} n{\cal R}^n = \frac{\cal R}{(1-{\cal R})^2}, 
\end{eqnarray}
and by Eq.~(\ref{eq:tradeoff_LR}), where we have $\overline{n} \le {\cal L}^{-1}({\cal L}^{-1}-1)$---i.e.,  for a single run of qRAM, it is possible to continue the reuse of $\ket{\psi_0}$, on average, less than ${\cal L}^{-1}\left( {\cal L}^{-1} - 1 \right)$. This implies that `the higher the learning efficiency or equivalently the oracle's reliability, the lower is the reusability of the state of the input data. Such a limited reusability may impose the requirement of a higher rate of qRAM query, to access ``big'' input data.

{\em Optimal usage of qRAM.}---We herein demonstrate that the usage of qRAM can be optimized by cycling the state $\ket{\psi_0}$ of the input data to the fundamental bound to saturate the tradeoff relation. Hence, we consider an exemplary protocol as described below. The oracle operation is described by
\begin{eqnarray}
\ket{\psi_0}\ket{0_\alpha} &\xrightarrow{\hat{O}(c)}& \displaystyle \sum_{\tau=0,1}\Big( \sqrt{\xi_\tau \lambda_{+}}\ket{X_\tau}\ket{c=\tau} \nonumber \\
&&~~ + \sqrt{\xi_\tau \lambda_{-}}\ket{X_{\tau \oplus 1}}\ket{c=\tau \oplus 1} \Big), 
\label{eq:oracle_op_w}
\end{eqnarray}
with the states of the correct $\ket{c=\tau}$ and incorrect answer $\ket{c=\tau \oplus 1}$. Subsequently, a measurement $M_\alpha$ is performed, yielding the oracle's answer with outcomes $m_\alpha \in \{0,1\}$. Given the measurement result $m_\alpha \in \{0,1\}$, the post-measurement states $\ket{\varphi_{m_\alpha}}$ defined in Eq.~(\ref{eq:Apsi0}) can be obtained~\cite{Ueda96}. The processes including the oracle and the subsequent measurement, $\hat{O}(c) + M_\alpha$, result in a specific form of remaining state $\ket{\varphi_{m_\alpha}}$, such that
\begin{eqnarray}
\ket{\psi_0}\ket{0_\alpha} \to
\left\{
\begin{array}{l}
\displaystyle \ket{\varphi_{0}} = \sqrt{\frac{\xi_0 \lambda_{+}}{P_0}} \ket{X_0} + \sqrt{\frac{\xi_1 \lambda_{-}}{P_0}} \ket{X_1}, \\
\displaystyle \ket{\varphi_{1}} = \sqrt{\frac{\xi_0 \lambda_{-}}{P_1}} \ket{X_0} + \sqrt{\frac{\xi_1 \lambda_{+}}{P_1}} \ket{X_1},
\end{array}
\right.
\label{eq:remaining_st}
\end{eqnarray}
where $P_0$ and $P_1$ are given in Eq.~(\ref{eq:P_c}) and denote the probabilities of getting $m_\alpha=0$ and $m_\alpha=1$, respectively. 

Subsequently, $\hat{U}_{m_\alpha}$ is applied on the state $\ket{\varphi_{m_\alpha}}$ and an ancillary state $\ket{0}_\beta$. The optimal $\hat{U}_{m_\alpha}$ can be chosen, according to the identified $m_\alpha$, to maximize the reusability ${\cal R}$. Here, we can select $\hat{U}_{m_\alpha}$ in the form of
\begin{widetext}
\begin{eqnarray}
\hat{U}_{m_\alpha}&=&\Big( \cos{\Theta} \left( \hat{\sigma}_x^{{m_\alpha}\oplus1} \otimes \hat{\openone}_N \right) + i (-1)^{{m_\alpha}\oplus1} \sin{\Theta} ~\hat{C}_{X_0, X_1} \Big) \Big( \hat{\openone}_2 \otimes \hat{R}(\Theta) \Big),
\label{eq:Umalpha}
\end{eqnarray}
\end{widetext}
where $\hat{C}_{X_0, X_1}=\hat{\openone}_2 \otimes \ket{X_0}\bra{X_0} + \hat{\sigma}_x \otimes \ket{X_1}\bra{X_1}$, and $\hat{R}(\Theta)=\ket{X_0}\bra{X_0} + e^{i\Theta}\ket{X_1}\bra{X_1}$. Here, $\hat{\openone}_d$ is the identity of $d$-dimensional Hilbert-space. For each case of $m_\alpha$, the state $\ket{\varphi_{m_\alpha}}$ undergoes the transformation with $\hat{U}_{m_\alpha}$ as
\begin{widetext}
\begin{eqnarray}
\ket{0_\beta}\ket{\varphi_0} &\xrightarrow{\hat{U}_{m_\alpha=0}}& -i \sqrt{\frac{\xi_0 \lambda_{+}}{P_0}}\sin{\Theta} \ket{0}\ket{X_0} + \sqrt{Q_0}\ket{1} \left( \sqrt{\frac{\xi_0 \lambda_{+}}{P_0 Q_0}}\cos{\Theta} \ket{X_0} + \sqrt{\frac{\xi_1 \lambda_{-}}{P_0 Q_0}}\ket{X_1} \right), \nonumber \\
\ket{0_\beta}\ket{\varphi_1} &\xrightarrow{\hat{U}_{m_\alpha=1}}& \sqrt{Q_1}\ket{0} \left( \sqrt{\frac{\xi_0\lambda_{-}}{P_1 Q_1}} \ket{X_0} + \sqrt{\frac{\xi_1\lambda_{+}}{P_1 Q_1}} \cos{\Theta}\ket{X_1} \right) + i \sqrt{\frac{\xi_1 \lambda_{+}}{P_1}}\sin{\Theta}  \ket{1}\ket{X_1},
\label{eq:st_aU}
\end{eqnarray}
\end{widetext}
where
\begin{eqnarray}
Q_0 &=& \frac{\xi_0 \lambda_{+} \cos^2{\Theta} + \xi_1 \lambda_{-}}{P_0}, \nonumber \\
Q_1 &=& \frac{\xi_0\lambda_{-} + \xi_1\lambda_{+}\cos^2{\Theta}}{P_1}.
\end{eqnarray}
Because $\Theta$ can be written in terms of ${\cal L}$, the optimal $\hat{U}_{m_\alpha}$ is determined depending on a given oracle's reliability ${\cal L}$. Here, if we set $\Theta = \arccos{\sqrt{\frac{\lambda_{-}}{\lambda_{+}}}}$, then $Q_j$ becomes $\frac{\lambda_{-}}{P_j}$ ($j=0,1$) and the transformations in Eq.~(\ref{eq:st_aU}) are rewritten as
\begin{eqnarray}
\ket{0_\beta}\ket{\varphi_0} &\xrightarrow{\hat{U}_{m_\alpha=0}}& -i \sqrt{\frac{\xi_0 {\cal L}}{P_0}} \ket{0}\ket{X_0} \nonumber \\
        && + \sqrt{\frac{\lambda_{-}}{P_0}} \ket{1}\underset{\text{reusable state $\ket{\psi_0}$}}{\underbrace{\left(\sum_{\tau=0,1} \sqrt{\xi_\tau}\ket{X_\tau}\right)}}, \nonumber \\
\ket{0_\beta}\ket{\varphi_1} &\xrightarrow{\hat{U}_{m_\alpha=1}}& i \sqrt{\frac{\xi_1 {\cal L}}{P_1}} \ket{1}\ket{X_1} \nonumber \\
       && + \sqrt{\frac{\lambda_{-}}{P_1}} \ket{0}\underset{\text{reusable state $\ket{\psi_0}$}}{\underbrace{\left(\sum_{\tau=0,1} \sqrt{\xi_\tau}\ket{X_\tau}\right)}}.
\label{eq:st_aU_op}
\end{eqnarray}

After a secondary measurement $M_\beta$ is performed on the first mode of Eq.~(\ref{eq:st_aU_op}), the probabilities of the cases when the results are consistent ($m_\beta = m_\alpha$) and inconsistent ($m_\beta \neq m_\alpha$) are obtained, respectively, as
\begin{eqnarray}
Q_{m_\beta = m_\alpha} = \frac{\xi_{m_\alpha} {\cal L}}{P_{m_\alpha}},~~~~Q_{m_\beta \neq m_\alpha}=\frac{\lambda_{-}}{P_{m_\alpha}}.
\label{eq:Q}
\end{eqnarray}
Then, it is inferred---observing Eq.~(\ref{eq:st_aU})---that the correct query output $\ket{X_{\tau}}$ can be extracted with the probability $Q_{m_\beta = m_\alpha}$, unless ${\cal L}=0$. In other words, we can confirm that the oracle's answer obtained in $M_\alpha$ is correct if it is consistent with the outcome of $M_\beta$, i.e., $m_\alpha = m_\beta$. The overall probability of attaining $\ket{X_{\tau}}$ is subsequently given as $\sum_{m_\alpha} Q_{m_\beta = m_\alpha}P_{m_\alpha} = 1 - {\cal R}={\cal L}$, satisfying the tradeoff relation. Meanwhile, for the case of inconsistent results, i.e., $m_\beta \neq m_\alpha$, one can recover the state $\ket{\psi_0}$ of the input data, that is, conclusively reusable. It is worth noting that the probability $Q_{m_\alpha \neq m_\beta}$ obtained in Eq.~(\ref{eq:Q}) is optimal, as described in Eq.~(\ref{eq:P_reuse_malpha}). Subsequently, the reusability can be calculated as ${\cal R} = \sum_{m_\alpha = 0,1} P_{m_\alpha} Q_{m_\beta \neq m_\alpha} = \sum_{m_\alpha = 0,1} \lambda_{-}= 1 - {\cal L}$, saturating the tradeoff relation in Eq.~(\ref{eq:tradeoff_LR}). Therefore, in principle, $\ket{\psi_0}$ is allowed to be cycled until the correct output $\ket{X_{\tau}}$ is extracted, by achieving the fundamental bound of the reusability. This indeed provides us an optimal process of query (in principle) without any additional qRAM queries caused by the incomplete oracle. 


{\em Remarks.}--- In summary, we have derived a tight tradeoff relation between the the reliability of the oracle and the reusability of the quantum state of the input data. It manifests the fundamental limit on the possibility of reusing a state, initialized as a superposition of the ``big'' input data for a single run of qRAM. The derived tradeoff relation indicated that the more reliable the oracle, the lower was the reusability. This would impose the additional usage of qRAM accessing ``big'' input data with the query failure. However, even with the limited reusability, the overall query process could be optimized by cycling the state initialized once. In particular, the optimized process was shown to saturate the fundamental upper bound of the reusability limited by the tradeoff relation. Remarkably, it was shown that the learner could, in principle, arrive at the correct answer without any additional qRAM queries caused by the incomplete oracle; for example, when the oracle produces incorrect answers, the quantum state of the input data could be recovered with post-processing to be used again for query. Such a process could be repeated until the correct answer is extracted. Thus, the complexity of qRAM query would not increase even with an unreliable oracle. These results will be crucial, since the low usage of qRAM is highly desirable in QML. We believe that our work will provide a fundamental and practical insight on the QML.

{\em Acknowledgements.}--- We are grateful to Jinhyoung Lee for the fruitful discussions. JB and AD are grateful to Marcin Wie\ifmmode \acute{s}\else \'{s}\fi{}niak, Wies\l{}aw Laskowski, and Marcin Paw\l{}owski. JB would like to thank Junghee Ryu and Nana Liu for the discussions. This research was implemented as a research project on quantum machine learning (No. 2018-104) by the ETRI affiliated research institute. JB acknowledge the support of the R\&D Convergence program of NST (National Research Council of Science and Technology) of Republic of Korea (No. CAP-18-08-KRISS).

\end{document}